\documentclass[aps,11pt,showpacs,superscriptaddress,amsfonts,amssymb,amsmath]{revtex4}

\usepackage{graphicx}

\begin{document}

\title{
A comparison between the YBCO discharge experiments by E.\ Podkletnov \\ and C.\ Poher,  and their theoretical interpretations}

\medskip

\author{G.\ Modanese \footnote{Email address: giovanni.modanese@unibz.it \\
To appear with linked references in Applied Physics Research, Dec. 2013 \  (\url{http://www.ccsenet.org/journal/index.php/apr})
}}

\affiliation{Free University of Bolzano, Faculty of Science and Technology, Bolzano, Italy \medskip}

\date{September 24, 2013}


\begin{abstract}

\medskip

Experimental results recently published by C.\ Poher provide independent evidence for the anomalous radiation emitted from YBCO electrodes under short, intense current pulses previously reported by E.\ Podkletnov. The generation conditions are somewhat different: lower applied voltage, longer duration of the pulses, absence of a discharge chamber. The microstructure of the emitter is also different in the two cases. While Podkletnov's radiation beam is collimated, Poher's beam is more or less diverging, depending on the emitter type. In coincidence with the radiation emission Poher measures a strong recoil of the emitter, with maximum momentum of the order of 1 kg$\cdot$m/s. We compare and discuss several details of the experiments and give a brief outline of the proposed theoretical explanations. We also report numerical simulations of the maximum electromagnetic recoil force on a Josephson junction, as a benchmark for a possible alternative explanation of the recoil.

\end{abstract}


 \maketitle

\section{Introduction}
\label{intro}

The experimental work by C.\ Poher, summarized in \cite{apr} and updated in \cite{web}, consists of a long series of trials (more than 6000 discharges through YBCO emitters), made over a period of about six years. Numerous different emitter configurations were tested, in different conditions of voltage, pulse energy and pulse duration. The descriptions and pictures available are quite detailed, and several cross-checks were performed. Some of the fabrication methods are patented \cite{pat}. These experiments cannot be strictly regarded as a replication of the work by Podkletnov and Modanese published in 2001 \cite{jltp}, also because they have a different purpose and start from a definite theoretical premise, which was not present in the case of Podkletnov. Nevertheless, the effects observed are partly the same as reported by Podkletnov, and Poher cites \cite{jltp}. We thus believe that Poher's work can be regarded as an independent confirmation. In this section we give a brief outline of the two experiments; a more detailed comparison will follow in later sections. Throughout the paper we shall use the acronyms EP for E.\ Podkletnov and CP for C.\ Poher, also meaning for brevity their respective devices. We will not repeat the content of the original articles, which are clearly written and freely accessible. For CP, we shall make reference mainly to the published article \cite{apr}, and occasionally to the other cited updates.

\begin{figure}
  \includegraphics[width=10cm,height=7cm]{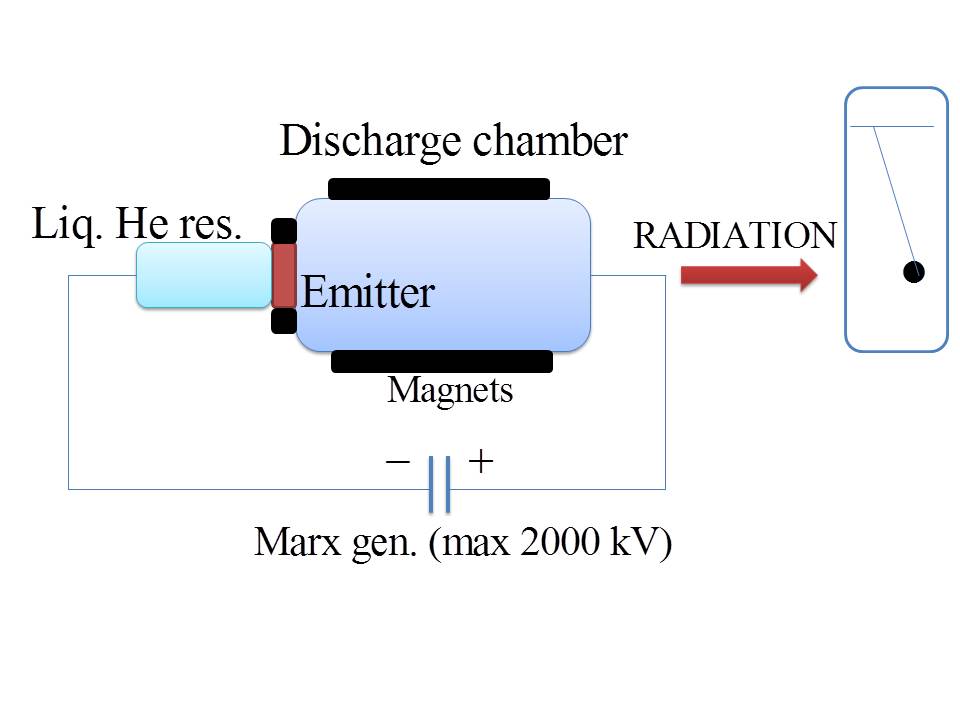}
\caption{Scheme of the device by E. Podkletnov (EP) for high-voltage discharges through a superconducting emitter. The emitter (red, diameter 10 cm) and the chamber (blue, diameter ca.\ 1 m) are surrounded by electromagnets. The emitter is cooled by lateral contact with a liquid helium reservoir. A Marx generator produces an over-damped high voltage pulse. The circuit has a distributed inductance of ca.\ 10 to 15 $\mu$H, but no load resistance. The emitted anomalous radiation propagates to a large distance in a collimated beam. Its effects are measured by the impact on small ballistic pendulums of variable mass and composition.}
\label{EP}       
\end{figure}

In both experiments a strong current pulse, of the order of 10 kA, is sent through a disc of YBCO high-Tc superconductor with diameter ranging from approximately 1 to 10 cm. The pulse is generated by the discharge of a capacitors bank. In EP, the total capacitance is of the order of 1 nF (Marx generator), and the capacitors are charged to several hundreds of kilovolts; in CP the capacitance is much larger (of the order of 100 $\mu$F) and the maximum voltage is of the order of some kilovolts. EP discharge circuit comprises a large vacuum chamber at very low pressure. The firing of the Marx generator triggers in this chamber a spark discharge, which works as a switch but probably also has some other, still unclear role in the process. In CP the pulse is switched by a solid-state thyristor. In both cases the anomalous effects are observed at a temperature well below the critical temperature of the superconducting emitters (90-92 K). CP has his emitters submerged in liquid nitrogen, while EP uses a liquid helium reservoir in contact with the emitter (which is usually larger and thicker and faces the vacuum chamber on one side, and for this reason is more difficult to cool down). The duration of the discharge is of the order of 1 $\mu$s for EP, with short risetime, and of 100 $\mu$s for CP (see Table 1). The emitter of EP is placed in a strong static magnetic field and is of the melt-textured type, i.e. with a globally ordered bulk crystal structure. CP tried emitters with various microscopic structures, but the most efficient for recoil production are those sintered with randomly oriented grains \cite{web}; these emitters generate an anomalous radiation with no focusing. Normal-conducting layers (EP, CP) or inter-grain boundaries (CP) are present in the emitters. These regions appear to be crucial for the production of the effects. As we shall see, their role is probably to host a strong electric field and possibly, in the EP case, to reduce the skin effect in the current fed to the emitter through the metallic electrode.

 \begin{figure}
  \includegraphics[width=10cm,height=7cm]{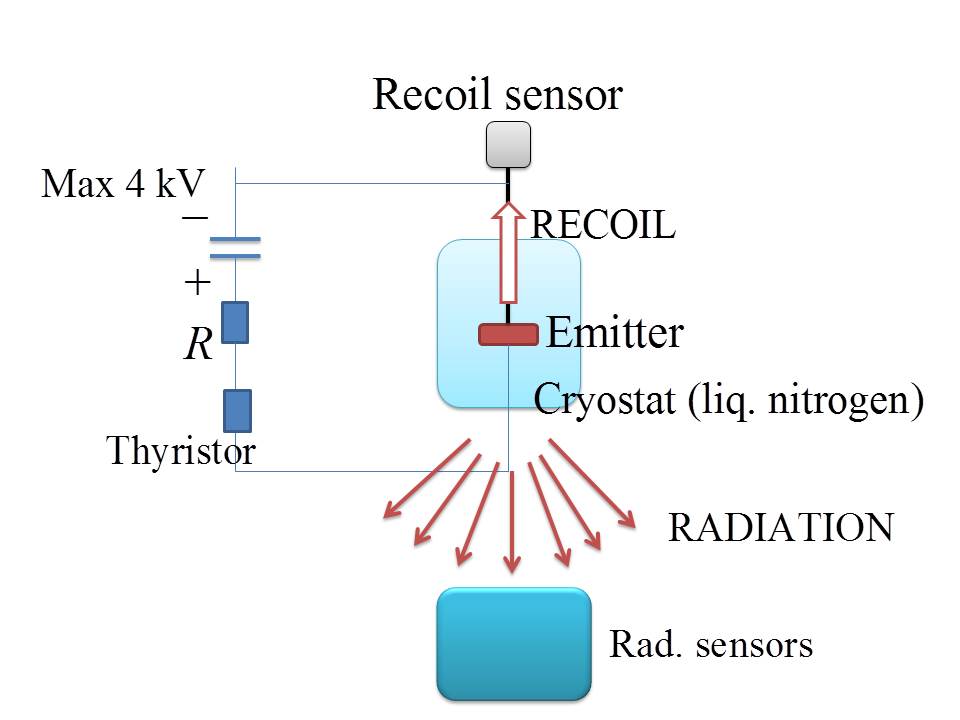}
\caption{Scheme of device by C. Poher (CP) for medium-voltage discharges in a superconducting emitter. The emitter (red, diameter 1 - 8 cm) is suspended in a liquid nitrogen bath and mechanically coupled to a recoil sensor. The discharge is produced by a capacitors bank and switched by a thyristor. There is a distributed inductance in the circuit of the order of 1 $\mu$H and a load resistance of the order of 0.1 $\Omega$. The emitted radiation is measured in a shielded box of sensors placed approx. 25 - 50 cm below the emitter and has an angular distribution which depends on the emitter type. A strong recoil is always detected in the opposite direction to the radiation.}
\label{CP}       
\end{figure}

Two kinds of anomalous effects are observed in coincidence with the discharges. EP reports the emission of a radiation pulse which propagates orthogonally to the cathode, towards the anode and beyond it, in a collimated beam, apparently without any attenuation even through thick metal plates, up to a distance of at least 150 m. The beam conveys to small free targets of any composition (ballistic pendulums with mass up to 50 g) a momentum proportional to their mass, imparting them a velocity of the order of 1 m/s, thus with a large instantaneous acceleration. CP also observes a similar acceleration beam, but only up to a distance of ca.\ 1 m and not so focused. The acceleration, estimated from the effect on small masses connected to a piezoelectric sensor, was of the order of 0.1 m/s$^2$ at low voltage \cite{apr} and much higher for higher voltage discharges, for instance 355 m/s$^2$ for a 1604 V discharge \cite{priv}. The presence of the beam was also detected with other sensors. The most relevant effect for CP, however, is a strong recoil of the emitter in the direction opposite to the electron flow and to the emission of the acceleration beam. The recoil momentum varies in a range from a few g$\cdot$m/s to a few kg$\cdot$m/s, depending on the features and energetic efficiency of the emitters, on their surface and on the electric energy of the discharge. The recoil momentum can be measured because the emitter is mounted on a movable support and coupled to a mechanical or electromagnetic sensor. The recoil is only present when a superconducting emitter is used. In the EP device, on the other hand, no recoil was noticed, possibly because the emitter was rigidly connected to a heavy structure composed of the discharge chamber, massive solenoids, liquid helium reservoirs etc.

The energy efficiency of the CP emitter can be estimated from the recoil energy (which is measured directly, except for possible mechanical losses) and from the electric energy $U_e=I V_e \cos \phi \Delta t$ delivered to the emitter, where $V_e$ is the voltage drop on the emitter, $\Delta t$ is the pulse duration, $\phi$ is the voltage/current phase. Note that $V_e$ is much smaller than the total voltage, because large voltage drops occur in the rest of the circuit. In general the measurement of the emitter voltage $V_e$ is not simple, due to parasitic inductances. CP reports energy efficiency ratios from a few percent to 30\% and higher. See details of the energy balance in Sect. \ref{ee-CP}. 

The computation of the energy efficiency of the EP emitter requires the knowledge of the energy of the radiation beam. It is not clear, however, which fraction of the beam energy is actually captured by the targets. Furthermore, the emitter voltage has not been measured by EP. For some indirect estimates, see Sect. \ref{ee-EP}.

For both EP and CP it is quite clear that any residual electromagnetic radiation cannot be responsible for the observed acceleration of far targets. In fact, the anomalous radiation conveys a momentum which is certainly not related to the carried energy by the usual dispersion relation $E=cp$, but is much larger. One can estimate, for instance, considering EP data for the 18.5 grams pendulum, that the kinetic energy associated to the observed acceleration is of the order of $10^{-3}$ J and the momentum is of the order of $10^{-2}$ kg$\cdot$m/s. If this momentum had to be imparted to the pendulum by radiation pressure, the energy needed in the beam would exceed the total energy available in the discharge ($\sim 10^{3}$ J).

In the EP experiment, the laser-like focalization of the radiation beam appears to signal the occurrence of a coherent stimulated emission process in the whole bulk of the emitter. Key factors enabling this stimulated emission could be the melt-textured structure and the high frequency components of the current pulse. The theoretical model mentioned in Sect. \ref{VFS} gives an estimate for the $A$ and $B$ coefficients of spontaneous and stimulated emission. According to the CP theoretical model (Sect. \ref{univ}), stimulated emission would also occur in the sintered emitters at the level of the single grains.

The article is organized as follows. In Section \ref{further} we continue our analysis and comparison of the two experiments, discussing first for both the energy efficiency (\ref{ee-CP}, \ref{ee-EP}). In Sect. \ref{vel-EP} we discuss some features of the EP experiment: propagation velocity of the radiation, features of the vacuum spark discharge, skin effect. In Sect. \ref{calib} we return to the CP experiment, focussing on the delicate issue of accelerometer calibration. In Sect. \ref{Re} we briefly touch upon the estimate of the dynamical resistance of the emitter. In Sect. \ref{convent} we discuss a possible conventional explanation of the recoil observed by CP, namely we estimate an upper limit on the mechanical effect of the non-linear self inductance of the emitter. We employ for this a simplified model, representing the emitter as made by intrinsic Josephson junctions described by the RSJ equations. The recoil momentum per unit mass estimated in this way turns out to be much smaller than the momentum really observed in the experiment, so this explanation does not appear to be viable. The situation may be different if one admits the presence of a strong electric field in the insulating or inter-grain layers of the emitter (Sect. \ref{convent2}). In Sect. \ref{theo} we review the new theoretical models proposed for the explanation of the experiments. They are mainly based on the concept of anomalous gravitational fluctuations (Sect. \ref{VFS}) or the universons theory (Sect. \ref{univ}). Further theoretical ideas are mentioned in Sect.\ \ref{other}. Sect. \ref{concl} presents our conclusions.

\begin{table}
\begin{tabular}{ccccccc}
\hline\noalign{\smallskip}
 & {\bf Pulse  }  
& {\bf Average } 
&  {\bf Max total}
& {\bf Emitter}
& {\bf Emitter}
& {\bf Target } \\
 & {\bf   duration} 
& {\bf   current} 
&  {\bf   voltage}
& {\bf   voltage} 
& {\bf   energy} 
& {\bf   energy}  \\
 &  $\Delta t$ (s) 
&  $I$ (A)
&  $V$ (kV)
&  $V_e$ (V)
&  $U_e$ (J)
&  $U_T$ (J) \\
\noalign{\smallskip}\hline\noalign{\smallskip}
{\bf EP} & $10^{-6}$ 
& $10^{3}$
& 2000
& 1 - 10 
& $10^{-2}$ 
& $10^{-2}$ 
 \\
 &  
& 
&
& (estim.)
&(estim.)
&(beam*) \\
\noalign{\smallskip}\hline\noalign{\smallskip}
{\bf CP} & $10^{-5}$-$10^{-4}$ 
& $10^{3}$-$10^4$
& 4.5
& 150 (max)
& $1$ - $20$ 
& $0.1$ - $10$ 
 \\
 &  
& 
&
& 
&
&(recoil) \\
\noalign{\smallskip}\hline
\end{tabular}
\caption{Magnitude order of some experimental parameters. For CP the data are mainly referred to the emitters described in \cite{apr}; more recent emitters have better performances. (*) Note that for EP the target energy of a single ballistic pendulum is of the order of $10^{-3}$ J; the total beam energy can only be guessed.}
\label{tab1}
\end{table}

\section{Further analysis of the experiments}
\label{further}

\subsection{Energy efficiency (CP)}
\label{ee-CP}

According to CP, the recoil momentum is always, to a good approximation, proportional to the total electric energy $U_{tot}$ of the discharge ($U_{tot}=CV^2/2$). For any emitter it is then straightforward to define a {\em total} efficiency coefficient, which gives the recoil momentum for unit of total energy. For emitters having the same structure, but different surfaces, the recoil momentum is found to be proportional to the surface, so a total efficiency coefficient per J and cm$^2$ can be also defined. Typical values for low-efficiency emitters described in \cite{apr} are 0.1 g$\cdot$(m/s)/(J$\cdot$cm$^2$).  Typical values for efficient multi-layer emitters are 100 times larger or more \cite{web}.

Another ``local'' energy efficiency coefficient of the emitter can be defined, as mentioned, as the ratio between the recoil energy and the sole electric energy $I V_e \cos \phi \Delta t$ available on the emitter. From the available data it is possible to estimate its magnitude order, but an exact evaluation would require further dedicated measurements. According to \cite{apr}, Sect. 3.1, the energy transferred to the EM3 emitters was 3 to 4\% of the stored energy. The same figure approximately applies to other emitters as well \cite{priv}. (In this connection, also note that the voltage readings of Fig.\ 5 of \cite{apr} are not reliable for an estimation of the emitter voltage.)

Consider for instance a 1604 V discharge of total energy 265.6 J into a quadruple type-V emitter \cite{priv}. This discharge imparted to the alternator a propelling momentum of 905 g$\cdot$m/s. Considering a total mass for the emitter of the order of 1 kg, the emitter energy efficiency turns out to be ca.\ 25\%. This might be underestimated, however, for the following reasons: (1) The energy of the recoil does not include the energy of the anomalous radiation. (2) When the recoil momentum is large, mechanical losses in the alternator can be relevant. 
(3) When the emitter undergoes a violent recoil, it often causes the expulsion of jets of liquid nitrogen from the cryostat. This phenomenon, which can be clearly seen in the video recordings, prevents a part of the recoil momentum from reaching the alternator. CP has demonstrated in several ways \cite{priv} that the recoil is the cause of the jets, and not the opposite. It is impossible, he has shown, to produce in liquid nitrogen heat and pressure levels sufficient for ``thermal propulsion'' through a microsecond current pulse.
(4) One may argue that the emitter could be made lighter by using different materials for the non-superconducting parts, while leaving the superconducting parts unchanged. Admitted the recoil momentum stays the same, in that case one would obtain a larger recoil energy.

\subsection{Energy efficiency (EP)}
\label{ee-EP}

The emitter voltage was not measured by EP, therefore it is difficult to compute the local energy efficiency. In his partial replication of the EP experiment \cite{Junker}, T. Junker made an attempt to measure it, but the strong disturbances due to the high voltage discharge spoiled the measurement. It is possible to do a theoretical calculation of the emitter voltage, based on a representation of the emitter as a series of intrinsic Josephson junctions. This is a good approximation for melt-textured YBCO, according to the classical work of Kleiner \cite{Kleiner}. The equations of the Resistively Shunted Junctions (RSJ) model (see also Sect.\ \ref{convent}) can be solved numerically when the emitter is coupled to an external oscillating circuit \cite{Nova} or receives an arbitrary current pulse. The simulation shows that there is an (almost linear) relation between the emitter voltage and the main frequency Fourier component of the pulse. For the frequency components of EP pulses, however, the estimated voltage turns out to be quite small, of the order of 1 V or less. This corresponds to an emitter energy of $10^{-3}$ J, which is also the magnitude order of the energy of a single target pendulum. The total beam energy is probably larger, so the theoretical model should be corrected to possibly include (1) some other junctions with large resistance (in addition to the intrinsic junctions), located for instance near the insulating layer; (2) higher frequency components of the current, besides the main component $1/\Delta t$, possibly generated in the vacuum discharge or due to the short rise time of the Marx generator. The values of $V_e$ given in Table I range between 1 and 10 V, thus comprising these two corrections.

\subsection{Propagation velocity, vacuum spark discharge, skin effect (EP)}
\label{vel-EP}

The propagation velocity of the anomalous radiation was measured by EP after the main experiment over a distance of 1211 m, using two synchronized rubidium clocks triggered by fast piezoelectric sensors \cite{Ch8}. The startling and still unconfirmed outcome was an apparent superluminal velocity of $(64 \pm 1)c$. Note that this is not in contrast with causality, as discussed in \cite{vel}, since the correlation between two sensors placed on the radiation path does not include the initial time for the generation of the pulse in the source. (The latter cannot be measured with a comparable accuracy.) Instead, the observed superluminal correlations appear in our opinion to confirm the quantum nature of the anomalous radiation phenomenon (compare Sect. \ref{VFS}); according to this interpretation, distant detections of the beam amount to measurements made on the same wavefunction.

A recent analysis of the EP discharges \cite{epr} in the vacuum chamber has evidenced a strong similarity to``vacuum spark discharges'' with superconducting cathodes. Accordingly, EP discharges would not involve chain ionization in a pre-existing gas, but evaporation of electrode material, as described for instance by \cite{Kuchler} for ``vacuum switches'' and by earlier articles on the vacuum spark discharge \cite{Korop,Koshelev}.

The same analysis also discusses the possible role of the normal layer in EP emitter in the suppression of the skin effect.
If there was a direct contact between a metal electrode and the superconducting emitter, the skin effect at high frequencies might prevent the current from passing through the bulk of the superconductor.  
Although no definitive conclusion on the internal skin effect in the emitter was reached, based on the available literature, it is believed that it is absent or moderate in YBCO in the $ab$ direction, for current flow in the $c$ direction. 
An important role of the normal layer in the EP emitter could then be that of avoiding direct metal-superconductor contact and re-distribute the current on the bulk before it is injected into the superconductor. 

The normal layer also hosts a strong electric field, which is supposed to be essential in the universons theoretical model (Sect. \ref{univ}). In Sect. \ref{VFS} we discuss whether this field could improve the spontaneous emission of anomalous radiation according to the gravitational vacuum fluctuations model.

\subsection{Calibration of the accelerometer (CP)}
\label{calib}

In \cite{apr}, Sect. 2.6.1, CP mentions a calibration procedure of the piezoelectric accelerometer based on dropping a small test mass from certain heights. This procedure was described in more details in \cite{priv} and is straightforward for forces in a low frequency band, but requires an extrapolation in order to be applied to the impulsive microsecond forces observed in the experiment. The figure of 8.8$\cdot 10^{-8}$ kg$\cdot$m/s given in Sect. 3.2.2 for the recoil momentum is not directly compatible with the low-frequency calibration constant of Fig. 8 (12.33 mV per Newton). An extrapolation to higher frequencies has been implicitly applied. Actually, what was measured was not the immediate response of the accelerometer to the force beam, but the residual oscillations of the test mass connected to the accelerometer after the action of the beam. During the pulse the output of the accelerometer is considerably affected by inductive disturbances, which are only partially screened by the aluminum enclosures. This time sequence is clearly seen from the oscilloscope readings. The inductive disturbances are very short-lived and erratic. On the contrary, the voltage output of the accelerometer during the residual oscillations is always proportional to the electric energy of the discharge, and in turn proportional to the output of the alternator (except when the recoil is so strong that some of the recoil momentum is dissipated before it can reach the alternator). Since the accelerometer measures the final velocity of a test mass placed in the force beam, it behaves in practice like the pendulums of EP, but in a kHz band instead of a Hz band. 

The effect of a short force pulse on a harmonic oscillator, with special reference to the balance between the transmitted energy and momentum and to the specific features of a piezoelectric sensor, has also been discussed in \cite{Ch8}.

\subsection{Dynamical resistance of the emitter}
\label{Re}

 The concept of ``dynamical'' resistance of the emitter plays an important role in both experiments.

CP: The resistance of the emitter during the discharge is usually different from the resistance measured with a four-contact ohmmeter before the discharge. For certain sintered emitters the two values can differ by magnitude orders, probably due to the presence of oxide layers. The dynamical resistance is estimated from the plot of the current vs time, as close as possible to the critical dampening condition.

EP: A puzzle remains concerning the origin of the apparent large resistance of the emitter to short pulses: the V/I ratio during the gas discharge was approximately 100 V/A, and the discharge looked overdamped. Several possible explanations were analysed: complex impedance of superconductors for high-frequency current, radiation reaction, resistance due to the intrinsic Josephson junctions for current flow in the c-axis of YBCO. None of these could explain the observed resistance. If the interpretation of the discharge as vacuum spark is correct (Sect.\ \ref{vel-EP}), that could offer an alternative explanation.

\section{Are there any conventional explanations for the emitter recoil?}
\label{convent}

Generally speaking, it is possible to generate electromagnetic vibrations or recoil forces in a device which has a mobile part and a certain self-inductance by feeding a variable current to the device. CP has made some trials replacing the superconducting emitter with a woofer, in order to test the reaction of his alternator when it is mechanically coupled to the mobile magnet of the woofer \cite{web}. He found that when an overdamped current pulse was sent to the woofer, the alternator signal was much weaker than the signal produced by superconducting emitters. Of course, the period of the woofer oscillations was also much longer, and the supplied current much smaller than with the superconducting emitters. In spite of these differences, one might wonder if the recoil of the superconducting emitter could be in part of electromagnetic origin, like that of the woofer. (We say ``in part'', because a part of the recoil momentum might be needed to balance the momentum carried by the anomalous radiation. The total amount of the radiation momentum has not been measured directly, as this would require a large-angle integrating sensor.) 

One immediate objection to the hypothesis of an electromagnetic recoil is that the superconducting emitters do not have any mobile parts. We can however regard as mobile part, in very simple terms, the superconducting electrons, which are free to move frictionless inside the superconductor and only feel a force (with some delay) at its border. In the following of this section we shall elaborate on this idea.

When electric current flows in an ohmic conductor, the total momentum imparted by the electric field to the electrons and to the positive ions averages to zero over practically any short time interval. Experimentally, thermal dissipation is observed, but no recoil of the conductor. This is due to the constant scattering of the accelerated electrons by the ions (accelerated in the opposite direction). The situation can be modeled theoretically either in the classical Drude theory or in the Fermi-gas quantum picture. A recoil effect might in principle be obtained in a hot plasma, if between successive scatterings the accelerated electrons emit electromagnetic radiation in one preferred direction (Fig.\ \ref{RSJ}). For a field strength of the order of $10^6$ V/m and a collision time of $10^{-9}$ s, the radiated energy is typically less than 1 part in $10^{20}$. The photon energy-momentum ratio $p=E/c$ implies that the ``imbalance'' momentum carried away by radiation is always extremely small. 

\subsection{Simulation of recoil forces on a Josephson junction}
\label{simulation}

\begin{figure}
  \includegraphics[width=10cm,height=7cm]{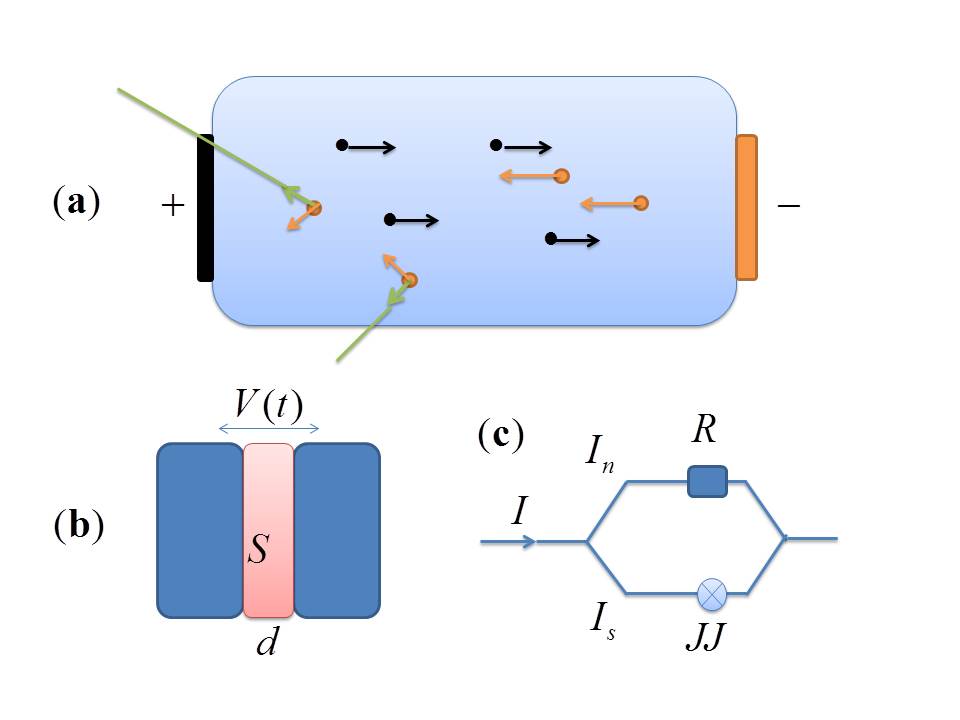}
\caption{(a) Mechanism for a very small recoil of a plasma during an electric discharge. Suppose that between subsequent scatterings the accelerated electrons (orange) emit photons (green) in a preferred direction. Then some momentum $p$ is lost from the plasma, of the order of $p=E/c$, where $E$ is the radiated energy (typically less than 1 part in $10^{20}$ of the electron energy). (b) Josephson junction of surface $S$, thickness $d$. A voltage $V$ (time-dependent, in general) is applied to the junction. (c) RSJ model (Resistively Shunted Junction): a real Josephson junction is equivalent to an ideal junction JJ obeying eq.s (\ref{RSJeqs}) in parallel to a resistor $R$. The total current fed to the junction splits in two parts, a normal electron current $I_n$ and a tunneling pair supercurrent $I_s$. Usually $I_n \ll I_s$. }
\label{RSJ}       
\end{figure}

In a superconducting material one can imagine a non-stationary process in which an electric field accelerates electron pairs and positive ions in opposite directions, thus creating a recoil that cannot immediately be balanced by ohmic friction. The balance is restored a bit later at the normal-superconducting junctions at the border of the material. The situation is difficult to model theoretically, also because the allowed electric field configurations must be compatible with some time-dependent evolution equation for the superconductor (compare also Sect. \ref{convent2}). A general upper estimate appears to be possible for the recoil of a Josephson junction described through the RSJ model (Fig. \ref{RSJ}). This estimate can then be applied to real cases, like sintered ceramic superconductors (with inter-grain junctions) or melt-textured ceramic superconductors (with intrinsic junctions, see \cite{Kleiner}). 

In the RSJ model a real Josephson junction is described as the parallel of an ideal junction and a resistance. The ideal junction carries only supercurrent and obeys the Josephson equations
\begin{equation}
\left\{ \begin{array}{l}
{I_s}(t) = {I_J}\sin \phi (t)\\
\phi '(t) = \frac{1}{{{\phi _0}}}V(t)
\end{array} \right.
\label{RSJeqs}
\end{equation}
where $I_s$ is the supercurrent, $I_J$ the critical current of the junction, $\phi$ is the Josephson phase, $\phi_0=2e/h$ and $V(t)$ is the applied voltage. $R$ is the resistance of the junction in its normal state and depends on the material, on the surface $S$ and on the thickness $d$. The general relation $I_J R=V_0$ also holds, where $V_0$ is a characteristic voltage which depends only on the material; for niobium junctions, for instance, one has $V_0$=2.2 mV. For YBCO junctions $V_0$ is usually in the range from 0.1 to 8 mV, depending on the micro-structure \cite{Waldram}. The simplest case is that of a junction subjected to a constant voltage; in that case $I_s$ oscillates in a pure sinusoidal way with frequency $\omega=V/\phi_0$. When the junction is inserted into an arbitrary external circuit, the voltage is not constant and $V(t)=RI_n(t)$. The simulation of the circuit requires to solve two coupled non-linear differential equations. It can be shown that real junctions in series become synchronized, so it is sufficient to consider only one junction \cite{Nova}.

\begin{figure}
  \includegraphics[width=15cm,height=10.5cm]{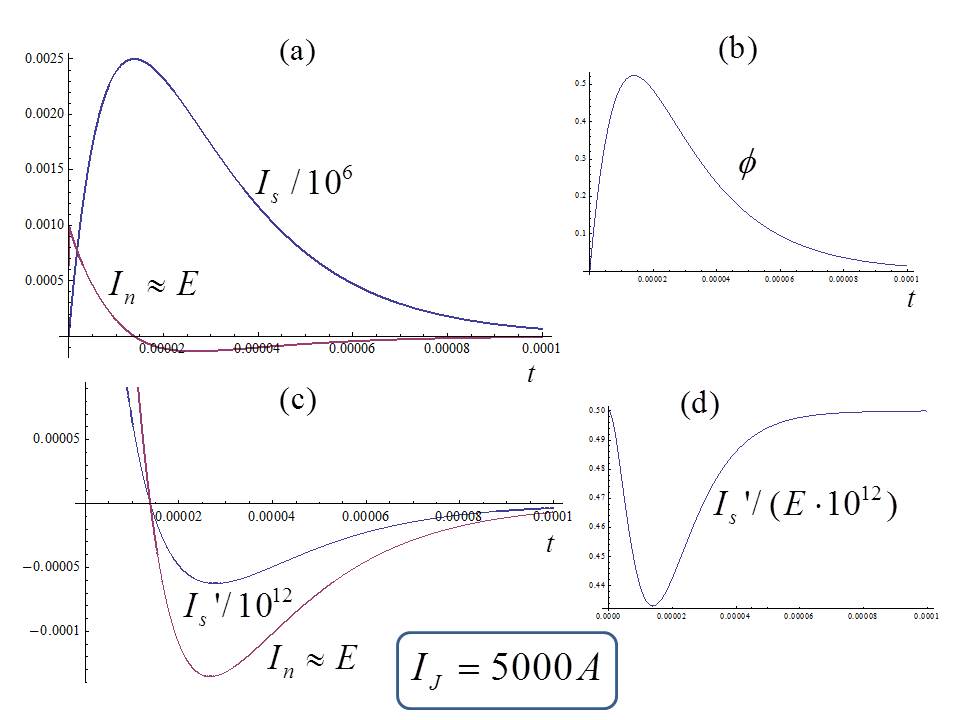}
\caption{Simulation of supercurrent, normal current and electric field in a Josephson junction biased with a microsecond pulse, according to eq.s (\ref{RSJeqs}). The normal resistance $R$ of the junction is $10^{-7}$ $\Omega$, its thickness $d=10^{-7}$ m, its critical current $I_J=5000$ A.  With these values, $E$ and $I_n$ have numerically the same intensity. Current and electric field are given in standard units (A, V/m). Time is in microseconds. (a) Supercurrent vs. normal current and electric field. (The total current is practically equal to the supercurrent.) Supercurrent and field are out of phase, showing the inductive behavior of the junction. (b) Time-dependence of the phase $\phi$. Note that it is very similar to the time-dependence of $I_s$. This is because $I_s$ is far from its limit value $I_J$; in these conditions $\sin \phi \sim \phi$. Compare Fig. \ref{2600A}. (c) Derivative of the supercurrent compared to normal current and electric field. A certain proportionality between $I'$ and $E$ is apparent (non-linear induction effect, electron acceleration) but not exact; the ratio varies between approx. 0.43$\cdot 10^{12}$ and 0.5$\cdot 10^{12}$ (d). }
\label{5000A}       
\end{figure}

\begin{figure}
  \includegraphics[width=15cm,height=10.5cm]{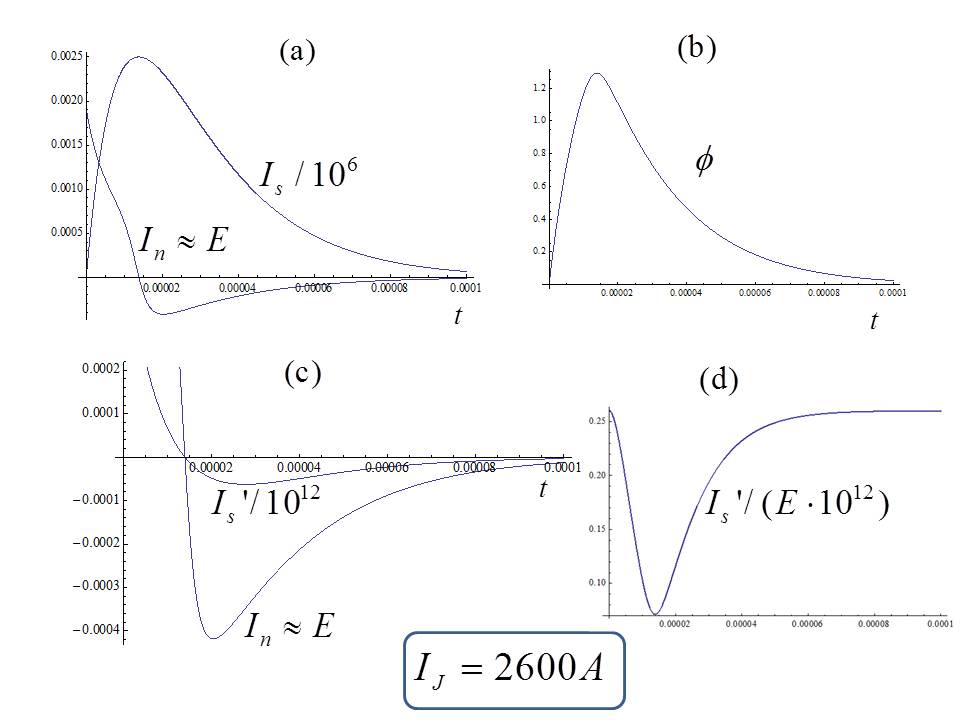}
\caption{Same as in Fig. \ref{5000A}, but with critical current $I_J=2600$ A.  The maximum current (ca. 2500 A) is now close to $I_J$. As a consequence, the Josephson phase $\phi$ is not as small as in the previous case and the factor $\sin \phi$ becomes more significant. A certain proportionality between $I'$ and $E$ is still apparent (non-linear induction effect) but less exact than in the previous case; the ratio varies between approx. 0.08$\cdot 10^{12}$ and 0.26$\cdot 10^{12}$ (d). }
\label{2600A}       
\end{figure}

Results are given in Fig.s \ref{5000A}, \ref{2600A}. Further details will be published elsewhere. For the purpose of the numerical simulation we can choose a value of $I_J$ which corresponds to that of the entire emitter. (In reality, the emitter contains many junctions in parallel and in series, but in the end we compute a force per unit mass, which is independent from the size of the junction.) Let us take an external pulse with a peak of ca. 2500 A. In the first simulation (Fig. \ref{5000A}) we chose a ``safe'' critical current $I_J=5000$ A. From the relation $I_J R=10^{-3}$ V this implies that $R=10^{-7}$ $\Omega$. In the second simulation (Fig. \ref{2600A}) we chose a critical current $I_J=2600$ A, which is only slightly larger than the total current. See the figure captions for further comments.

The delay between the normal current and the supercurrent shown by the simulations is important, because it could give rise to a momentary imbalance between the opposite forces exerted by the field on the positive ions and on the electrons. In a certain sense the junction behaves like a non-linear inductor. This property of the Josephson junctions is well known \cite{Waldram,Nori}. It can be seen from the figures that the ratio between the electric field and the derivative of the current is not exactly constant, especially when the supercurrent is close to the critical current.

The simulations show that the peak of the normal current is of the order of $10^{-4}$ A. At that moment, the voltage on the junction is therefore of the order of $10^{-11}$ V. The electric field $V/d$ is of the order of $10^{-4}$ V/m. The number of positive ions in the junction is $nN_A$, where $n$ is the number of moles; $n=1000m/\mu$, where $m$ is the mass in kg and $\mu$ the mass of one mole in atomic units. Let us consider for instance a niobium junction, with $\mu=92$. The total force on the positive ions of the junction is obtained by multiplying the field by the total positive charge. Finally we divide by the mass of the junction and obtain
\begin{equation}
a = \frac{F}{m} = \frac{{en{N_A}}}{m}E   \approx {\rm{1}}{{\rm{0}}^2}{\rm{\  m/}}{{\rm{s}}^2}
\label{accel}
\end{equation}

This acceleration is quite large, but it acts for a very short time. Furthermore, this is only an upper limit. The real final effect of this acceleration on the electrons and ions of the lattice depends on a unknown restoring force which is not present in the RSJ model, but must  be present in a real situation, probably arising at the normal-superconducting junctions where the external current flows into the superconductor. The delay between the restoring force and the force computed above defines the amplitude of the oscillation. (This delay is practically zero in an ohmic conductor.) 

Also note that the acceleration computed in eq.\ (\ref{accel}) is only referred to the mass of the Josephson junction, which can be at most a few percent of the total mass of the superconductor. The total acceleration will therefore be much smaller.

In conclusion, the magnitude order of the force due to the non-linear inductance of the superconductor does not seem sufficient to explain the observed recoil. 

\subsection{Electric field at superconducting/normal junctions}
\label{convent2}

\begin{figure}
  \includegraphics[width=10cm,height=7cm]{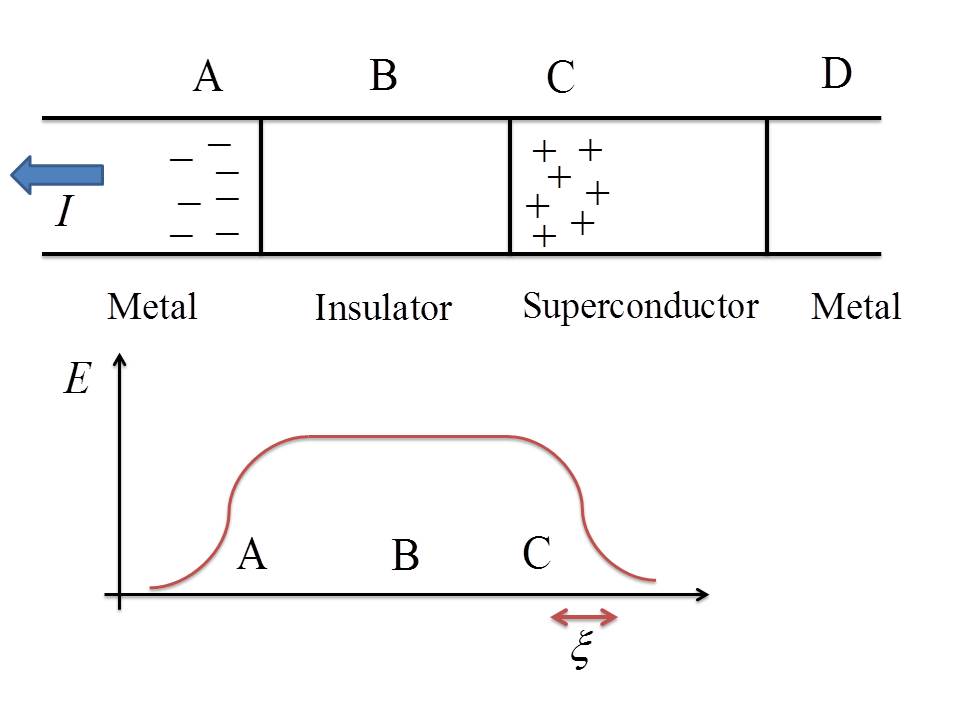}
\caption{Figure 6. Electric field in the insulating layer and depletion regions.
(Top) When a current pulse supplied through metal electrodes crosses an emitter composed of an insulator or poorly conducting layer and of a superconducting layer, negative charges tend to accumulate for a short time at the negative electrode, and to be depleted from the side of the superconductor which is in contact with the insulator. (Bottom) The electric field due to this charge inbalance is mainly located in the insulator, but also penetrates for a length $\xi$ into the superconductor.
}
\label{depletion}       
\end{figure}

The Josephson junction model employed in the previous section is useful because it allows an illustration of some principles and a quantitative evaluation in a special case, but insulating layers or inter-grain layers are often too thick to be regarded as Josephson junctions.
When a superconductor is in contact with an insulating or poorly conducting layer, electric fields can penetrate from the layer into the superconductor for a short characteristic length $\xi$ \cite{Waldram,Waldram2}.
This is one of the hypotheses of the theoretical interpretation of the anomalous emission proposed by CP (Sect. \ref{univ}). According to CP, the penetrating field strength can be of the order of $10^6$ V/m, causing an electron acceleration of the order of $10^{17}$ m/s$^2$. The experiments and theory of ref.s \cite{Waldram,Waldram2} cannot confirm this figure, because they refer to stationary situations where only small currents are present. The discharges in the EP and CP experiments appear instead to involve transient and more violent processes. Still we think that during the discharge the situation can be qualitatively represented as follows (Fig.\ \ref{depletion}). Because of the poor conductivity in the insulating layer B, an excess of electrons is temporarily produced in the metal electrode A, while the superelectrons are partially depleted from C. In this way, a strong electric field is generated in the insulating layer. The thickness of the partial depletion region C is of the order of $\xi$. In this region, the residual electric field accelerates the superconducting pairs in the opposite direction of the lattice ions. This temporary momentum imbalance is compensated when the superconducting pairs reach the other metallic electrode D and the supercurrent is again converted to normal current. In the region C there is therefore a transient force on the lattice, which can be much larger than the force in a Josephson junction, since the field is much stronger ($10^6$ V/m, compared to $10^{-4}$ V/m). This process can give rise to a recoil, however, only if the force acts for a sufficiently long time; we cannot presently estimate this time. Also recall that the recoil force observed in the CP experiment is proportional to the surface of the emitter and to the total electric energy of the discharge; it is not clear yet if transient recoil forces originating from temporary internal momentum imbalance would share these features. 

\section{Theoretical models}
\label{theo}

\subsection{Gravitational vacuum fluctuations}
\label{VFS}

\begin{figure}
  \includegraphics[width=11cm,height=8cm]{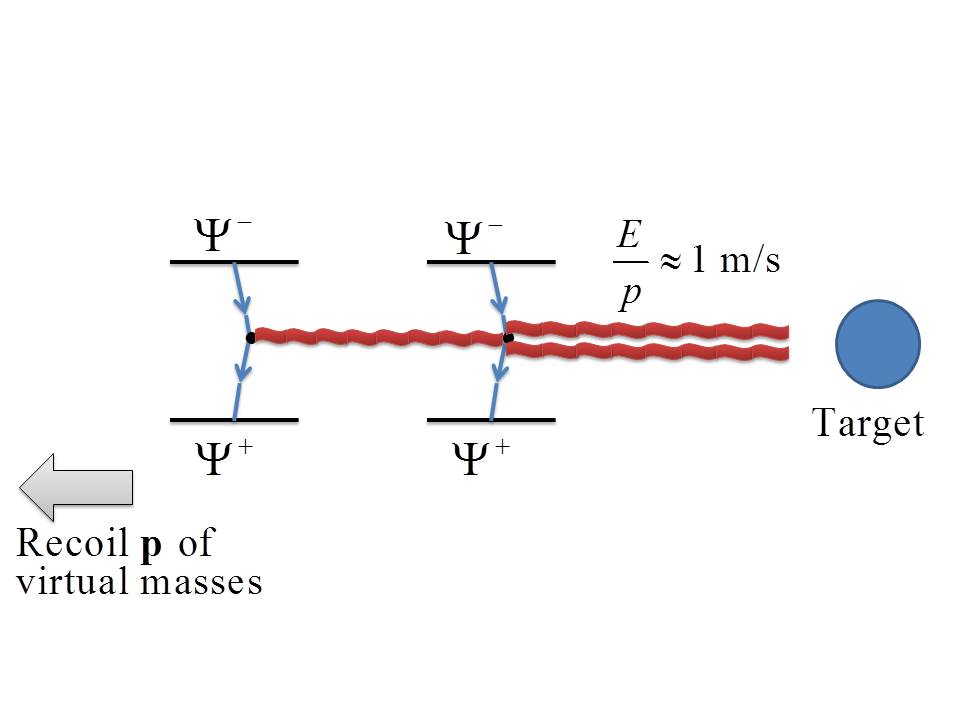}
\caption{Gravitational vacuum fluctuations model of the emission of anomalous radiation by superconductors. The high-frequency interaction of the superconductor  (not shown) with the gravitational vacuum fluctuations excites some of them from their symmetrical virtual mass pair state $\Psi^+$ to the corresponding antisymmetric state $\Psi^-$. In the subsequent decay, strongly off-shell virtual gravitons are emitted, which cause further stimulated emission in the bulk and propagate to the target. 
}
\label{vfs}       
\end{figure}

According to this model \cite{VF1,VF2}, the anomalous radiation beam is composed of virtual gravitons which are emitted by certain strong vacuum fluctuations of the gravitational field. These fluctuations, which are normally in a symmetric ground state $\Psi^+$, can be locally brought to an antisymmetric excited state $\Psi^-$ by the interaction of the vacuum with a superconductor, or more generally with coherent matter. The emission occurs when the fluctuations return to the ground state (Fig. \ref{vfs}). The fluctuations can be described as pairs of equal virtual masses which form bound states; the energy of their first excited state depends on the mass, therefore the virtual masses actually involved in the process are selected by the frequency of the excitation mechanism available. The whole process is virtual, which means that the fluctuations and their bound states exist only for a short time and that the emitted virtual gravitons cannot propagate to infinity, but only to a target. The virtual gravitons are, in quantum field theory jargon, completely ``off-shell'', and their $E/p$ ratio is very different from that of real massless particles. This corresponds to what is observed in the experiments: the anomalous radiation carries a very large momentum, in comparison to its energy.

In this model the conservation of total momentum after the emission of the virtual gravitons is guaranteed by the recoil of the virtual masses. Being quite large, the virtual masses can balance with their recoil practically any $E/p$ ratio required for the virtual graviton. ``Required'' is meant here in the sense that energy-momentum conservation in the final absorption of gravitons in the target also imposes a constraint, depending on the features of the target. The quantum nature of the whole process allows to explain how the emission recoil can be determined by the absorption target \cite{vel}.

The Einstein coefficients $A$ and $B$ for spontaneous and stimulated emission in the transition $\Psi^- \to \Psi^+$ are given, according to this model, by 
\begin{equation}
A = \left( {B\frac{{8\pi \hbar }}{{{\lambda ^3}}}} \right) = \left( {\frac{G}{{{\hbar ^2}}}|\langle {\bf{\hat d}}\rangle {|^2}\frac{{8\pi \hbar }}{{{\lambda ^3}}}} \right)
\end{equation}
where $\lambda$ is the wavelength of the emitted graviton and ${\langle {\bf{\hat d}}\rangle }$ is a matrix element involving the two virtual masses:
\begin{equation}
\begin{array}{l}
{\langle {\bf{\hat d}}\rangle } =
\langle {\psi ^ + }|{\bf{\hat d}}|{\psi ^ - }\rangle  = \frac{1}{{\sqrt 2 }}\left( {\langle 1| + \langle 2|} \right)\;{\bf{\hat d}}\;\frac{1}{{\sqrt 2 }}\left( {\left| 1 \right\rangle  - \left| 2 \right\rangle } \right) = \\
 = \frac{1}{2}\left( {\langle 1| + \langle 2|} \right)\left( {{m_1}{{\bf{r}}_1}|1\rangle  - {m_2}{{\bf{r}}_2}|2\rangle } \right) = \frac{1}{2}m{\bf{r}}
\end{array}
\end{equation}
Here ${{\bf{r}}_1} = \frac{1}{2}{\bf{r}}$, ${{\bf{r}}_2} =  - \frac{1}{2}{\bf{r}}$ and ${\bf r}$  is the displacement between the two virtual masses.

Numerical estimates of $A$ and $B$ are consistent with a single-pass laser action in the emitter. The mean free path for stimulated emission is of the order of $10^{-5}$ m and this leads to a large amplification factor already in a thickness of a few millimeters. A full rate equation has not yet been worked out, but an attempt was made to adapt the Frantz-Nodvik equation to this case \cite{Nova} and the condition for population inversion was evaluated in \cite{VF1}. 

The focalization of the beam observed in the EP experiment appears to be due to stimulated emission occurring over the bulk of the emitter, in direction orthogonal to the $ab$ planes of the superconductor. The versus of the emission appears to be related, in this picture, to the presence of a strong electric field in the normal layer of the emitter, but the exact role of the field in the emission of the anomalous radiation is still unclear. The electric field does not affect the $A$, $B$ coefficients, but it does influence the rate of the ``pumping'' transition $\Psi^+ \to \Psi^-$, because it contributes to the local vacuum energy density through its Lagrangian density $L_E=\frac{1}{2}\varepsilon_0 {\bf E}^2$; for strong fields, this contribution is comparable to that of the superconductor.

The recoil momentum acquired by the virtual masses can in principle be passed to matter through scattering. The classical scattering cross-section $\sigma$ for the impact of virtual masses on nucleons turns out to be very small. Again, the relevant interaction is probably that with coherent matter, but the amplitude of this interaction is still unknown. Therefore the model does not give, at present, a description of the recoil phenomenon.
 
\subsection{Universons theory}
\label{univ}

The theoretical interpretation proposed by CP for his experimental results is quite unconventional and is not based on existing models. Actually, this theory served as an initial stimulus for the experiment and as an effective guidance in its planning and execution. The theoretical model by CP is not contained in previous publications; it is exposed in the appendix of \cite{apr} and in some informal documents in the website \cite{web}. A paper by CP and P. Marquet \cite{Marquet} proves the compatibility of the model with General Relativity, but does not question its foundations. 

\begin{figure}
  \includegraphics[width=11cm,height=8cm]{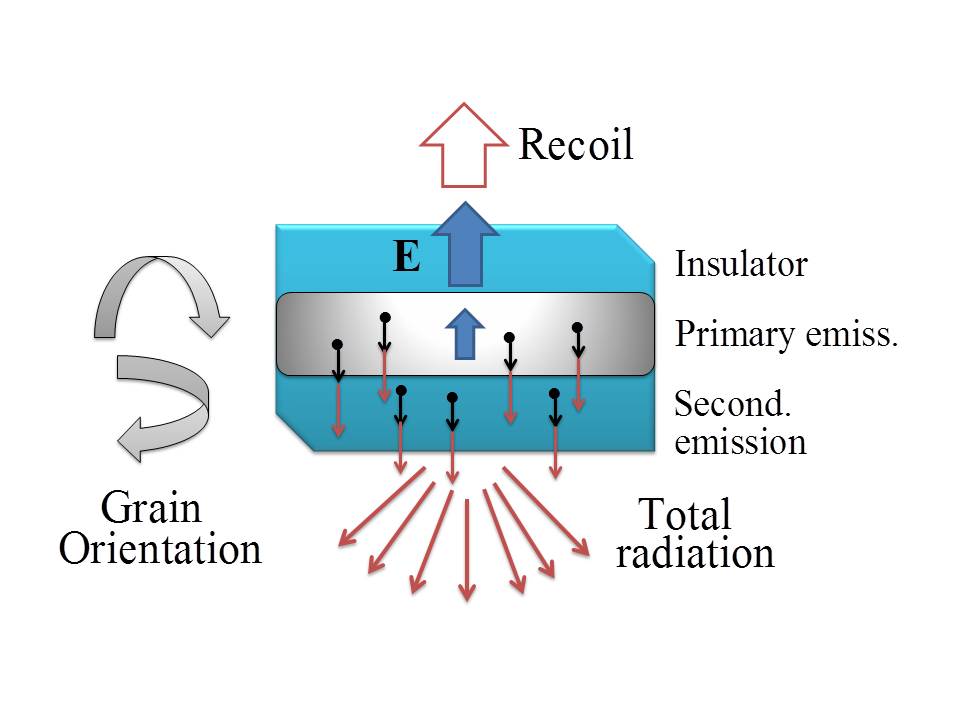}
\caption{Pictorial representation of the microscopic mechanism for the  emitter recoil and for generation of anomalous radiation, according to the universons theory by CP. In each single YBCO grain, the partial penetration of the electric field from insulating layers nearby accelerates the superconducting electrons (black) and make them emit universons (red). The variable orientations of the grains lead to a diverging total radiation. The whole process also involves a cosmic universons background (not depicted), which affects the total energy-momentum balance.}
\label{thCP}       
\end{figure}

The model postulates the existence in all space of a homogeneous and isotropic flux of particles called ``universons'' which propagate at light speed and are continually absorbed and re-emitted in a very short time by all massive elementary particles, thus giving rise to their inertia. (In this sense, the model has elements in common with theories like those of Puthoff-Haisch-Rueda, which explain the particles' inertia as a result of their interaction with electromagnetic vacuum fluctuations \cite{Haisch}.) The initial aim of the model was to address some problems in astrophysics, in particular the anomalies in the galactic rotation velocity curves. In this context the universons hypothesis is a substitute (more simple and natural, according to CP) for other kinds of dark matter.

From the point of view of the universons model, the superconducting emitter employed in the experiment interacts strongly with the natural universons background flux, and is able to divert a small part of the flux and extract some energy from it. This interpretation allows to circumvent a considerable conceptual difficulty, namely to explain how the observed anomalous radiation can convey to the targets a momentum which is much larger than the radiation pressure momentum $p=E/c$. In the universons model, the ``transmission balance'' is not limited to emitter and radiation, but also encompasses the surrounding background. Roughly speaking, a kind of vacuum pressure is involved, which is able to transfer much more momentum than could be carried by single particles. In this picture, the universons of the anomalous beam are not just absorbed in the target, but absorbed and quickly re-emitted.

As mentioned in Sect.\ \ref{ee-CP}, CP measurements show a proportionality between the recoil momentum and the electric energy of the discharge. Since the recoil energy is in turn proportional to the square of the momentum, an extrapolation of this proportionality relation to higher energies would imply that at some point the recoil energy exceeds the electric energy. CP has estimated in which conditions this should happen, supposed the proportionality relation keeps true. These ``excess energy conditions'' are quite far from the present experimental parameters. The presence of an excess output energy would clearly be a confirmation of the universons model. The vacuum fluctuations model of Sect.\ \ref{VFS} predicts the exchange of some momentum with the vacuum, but not an energy gain.

CP has proven within his model that an accelerated particle emits universons in an anisotropic way. The main idea behind the experiment is that during the discharge a large number of electrons in the emitter are strongly accelerated and emit universons in the direction of their acceleration. The acceleration would be possible thanks to the partial penetration of the electric field in certain regions of the superconductor, mainly near the borders of the YBCO grains and near the insulator-superconductor junctions. The product of large electron acceleration (up to $10^{17}$ m/s$^2$) and high electron density achieved in these conditions would be much larger than the analogous quantity obtained in other devices like X-ray tubes or particles accelerators. Moreover, the emitted universons could themselves accelerate further electrons in the bulk of the superconductor, thus leading to an amplification of the process. The fact that the many individual anisotropic fluxes of universons simultaneously emitted by all the atomic nuclei accelerations do not cancel the effect of the anisotropic fluxes emitted by all the primary accelerated electrons would be peculiar of superconductors.

\subsection{Other theoretical models}
\label{other}

According to N. Wu \cite{wu}, anomalous gravitational emissions in superconductors could be explained by a generalized gauge theory of gravitation that she developed over several years \cite{wu1}. 

In a series of papers \cite{rob}, G.A. Robertson looks at special features of superconductors for guidance toward the understanding of what properties could lead to anomalous gravitational phenomena. This is accomplished by first looking at properties for rapid power flow internal to the superconductor, applying these properties to energy radiated in gravitational wave, establishing an experiment to setup this power flow (which happens to look much like the EP experiment) and evaluating the theoretical calculations to the EP experiment. The theoretical calculations and the experimental data are shown to be surprising similar both in graphical form and value. 

It has been known for a long time \cite{halpern} that the probability of spontaneous or stimulated gravitational emission from atomic systems is so small that laser action can not be obtained in any realistic conditions. This conclusion is inevitable in the weak-field approximation, and if one considers only incoherent matter and real radiation with energy-momentum ratio $E/p=c$. Recently G.\ Fontana \cite{fontana} has studied the possibility of generating gravitational laser action in certain special circumstances, which can only occur at the junctions of two different superconductors, namely (a) transitions of Cooper pairs between s and d states, (b) super-radiance due to the macroscopic coherence of the superconductors, (c) suppression of the competing electromagnetic transitions due to the peculiar electrodynamics of superconductors.

In an unpublished report \cite{lewis}, R.A. Lewis analyses an early version of the CP experiment and the viability of an explanation of the observed phenomena through a vacuum field of the universon type. He observes that a phenomenological model involving
Dirac string flux tubes may summarize some features of the experimental data. He hypothesizes that vaporization of liquid nitrogen could contribute to the observed recoil. (This hypothesis has been however confuted by the most recent data of C. Poher with high-efficiency emitters.)

Further speculative works on EP and CP experiments have been published by Consiglio \cite{consiglio}, Sukenik and Sima \cite{su-si}, LaViolette \cite{laviolette}.

\section{Conclusions}
\label{concl}

We have analysed similarities and differences between the high-voltage discharge experiments by E. Podkletnov (EP) and C. Poher (CP). Both authors report the emission from YBCO electrodes of an anomalous radiation with small energy-momentum ratio, under conditions which are peculiar but not prohibitive. CP expands considerably the range of possible choices of emitter structure, applied voltage, duration and form of the discharge. He further measures a recoil momentum of the emitter which is always proportional to the total electric energy of the discharge; typical values for efficient emitters are of the order of 1 kg$\cdot$m/s per 200 J of electric energy. 

The recoil momentum is always opposite to the radiation momentum, but the exact relation between the two is tricky. On one hand, EP does not observe any recoil. This might be because the recoil is damped by a rigid structure (possibly the large discharge vacuum chamber, which is the main difference between the EP and CP devices). On the other hand, since the momentum imparted by the radiation on a target is proportional to the mass of the target, it is not obvious that one can detect the same momentum per unit target surface, no matter how many detectors one places, and no matter how much they weigh. We might call this the ``paradox of the virtual radiation'' (a term borrowed from the theoretical model of the effect based on vacuum fluctuations and virtual gravitons). It would imply that the recoil of the emitter depends to some extent on the target. There would also exist a maximum target mass, such that for larger masses the target acceleration would not be constant, but would decrease and tend to zero \cite{Ch8}.

A totally different view of the whole process is offered by the universons model. In that model, the background universons flux enters the local and total balance of energy and momentum and can completely alter it. For comparison, according to the virtual gravitons model there is in the emitter an exchange of momentum with the vacuum, but not an exchange of energy; and the propagation of the radiation and its interaction with the targets do not involve any further interaction with the vacuum.

Our theoretical simulation of the electromagnetic recoil of a Josephson junction suggests that that kind of recoil is much smaller than the one observed. More realistic simulations should take into account the penetration of strong pulsed electric fields into the insulating layers between the YBCO grains, and effects at the YBCO-metal junctions. This appears to be out of reach of the standard superconductivity theory, and will require further investigation.

Independently from the theoretical interpretations, the possible practical applications of both the anomalous radiation and the recoil phenomenon appear to be quite relevant. A working phenomenological model would be important, of course, in order to optimise the effects and to connect them to prior art. But the reported phenomena appear so new and peculiar that much further experimental and theoretical work will probably be necessary for that purpose.

\begin{acknowledgements}

I am grateful to C.\ Poher and D.\ Poher for experimental demonstrations and for very detailed discussions.

\end{acknowledgements}

\end{document}